\documentclass{iopart}
\usepackage{iopams}  
\usepackage{citesort} 
\usepackage{epsfig}

\newcommand{\be}{\begin{equation}}
\newcommand{\ee}{\end{equation}}

\newcommand{\sub}[1]{_{\mbox{\scriptsize #1}}}
\renewcommand{\vec}[1]{\bi{#1}}

\begin{document}

\article{Article}{A white-light trap for Bose-Einstein condensates}

\author{CA Sackett and B Deissler}

\address{Physics Department, University of Virginia, Charlottesville, VA 22904}

\ead{sackett@virginia.edu}

\begin{abstract}

We propose a novel method for trapping Bose-condensed atoms using 
a white-light interference fringe.  Confinement frequencies
of tens of kHz can be achieved in conjunction with trap depths
of only a few $\mu$K.  We estimate that lifetimes on the order of 10 s can
be achieved for small numbers of atoms.
The tight confinement and shallow depth permit tunneling processes to
be used for studying interaction effects and for applications in 
quantum information.

\end{abstract}


\section{Introduction}

The growth of interest in quantum computation has spurred 
investigations of few-particle interaction effects in a variety of
systems.  Bose-condensed atomic gases are no exception, and are in
fact a natural candidate for quantum 
information studies \cite{Calarco00}, since they
are good sources of particles in a completely
well-defined quantum state.   
However, because interactions in a
condensate are weak, they typically can only be observed through
mean-field effects with large numbers of atoms.
In contrast, few-particle phenomena have been observed
in several experiments using condensates
confined in optical lattices \cite{Orzel01,Greiner02}.  
These experiments rely on confining 
atoms relatively tightly within a lattice site to maximize 
interaction strength, and also maintain a relatively shallow
lattice depth so that tunneling processes can provide sensitive
dependence on the weak interaction energy.  These two requirements,
tight confinement and shallow depth, suggest that an important parameter
is the spatial size of the trap.  Optical lattices work for observing
effects like the Mott-insulator transition \cite{Jaksch98} because the size of
their lattice sites, typically about 500 nm, is sufficiently small.

Although atoms trapped in an optical lattice are interesting,
they are not ideally suited for quantum computing and related studies 
because a large number of sites are typically occupied, with 
spatial inhomogeneities making each site different and the close spacing
making individual addressing difficult.  We propose here a novel
trapping method that solves these problems by 
providing a confining potential similar to that of
one single site in an optical lattice.  This should allow detailed study of
interaction effects, perhaps controllable at the single-particle level.
The trapping mechanism is similar to that of
optical lattices, but instead of 
using a laser beam to generate a standing wave, we propose to 
use a broadband source to generate a white-light interference fringe.

A similar confining potential can be more conventionally obtained 
in a dipole-force trap generated by a single tightly focused laser beam.
At CNRS \cite{Schlosser01}, at trap of this type has been demonstrated with a
beam waist of about 1 $\mu$m, comparable to the confinement size
proposed here.
However, obtaining such a tight beam
requires complex focusing optics with high numerical aperture, which is
is expensive in terms of both money and optical access. 
In contrast, the white-light trap can use optics
of modest numerical aperture with substantial aberrations.
Also, the dipole trap is relatively weak in the direction along the beam axis,
while the white-light trap can give tight confinement in one, two, or 
three dimensions as desired.  However, the dipole force approach can provide
much stronger trapping potentials, can more readily provide and manipulate
multiple traps, and relies on more familiar laser physics.
Thus both approaches have unique features which could prove useful.

\section{White-Light Trap}

The principle of white-light interference is well known \cite{Born99}.
In an arrangement
such as in figure~1, there exists a plane $z = 0$ 
in which the optical path lengths
from the source are identical for both possible paths.  So, regardless
of the frequency of the light $\omega$, an interference maximum or minimum
will occur in this plane, with the sign of the interference depending
on the number of mirror reflections occurring in the two arms.
However, if the source emits a range of frequencies $\Delta\omega$,
then at distances $z \gg c/\Delta\omega$ the interference
maxima and minima for different frequencies will be uncorrelated and
the total intensity will be nearly constant.  
If $\Delta\omega$ is sufficiently large, the interference will
extend over only one oscillation and only a single trapping site results.  

\begin{figure}
\hfill\epsfig{file=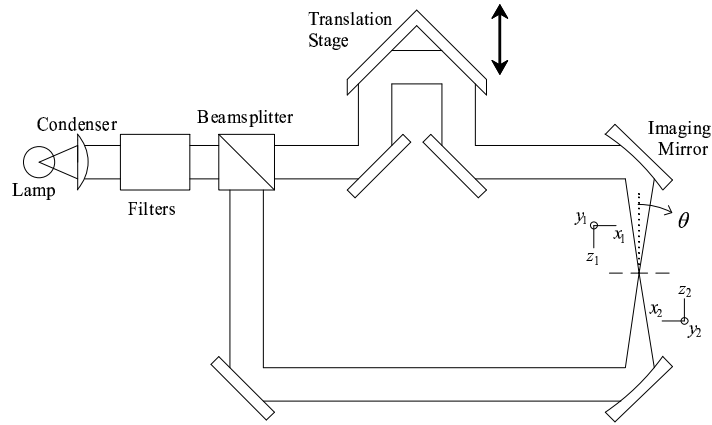,clip=,width=4in}
\caption{
Layout for white-light trap.  Light from a lamp is collected by a condenser
lens, filtered appropriately, and divided by a beamsplitter.
The resulting beams are directed counter-propagating and focused
to achieve high intensity.  The angle 
$\theta$ parametrizes the focusing. 
Interference occurs in the plane where the 
distance to the source along both paths is equal, indicated by the
dashed line.  The location of this plane can be adjusted using
translating mirrors, as shown.  The two small sets of axes represent
the coordinate systems used to calculate the interference term
in equation (\protect\ref{eq-mu12}).
}
\end{figure}

White light has not previously been used for atom trapping because 
the available light intensity permits only relatively weak traps.
However, a condensate provides a source
of atoms in the absolute ground state of their potential, so
the white-light trap can work
as long as sufficient intensity can be generated to provide at least
one bound state.
The brightest commercially available broadband source is a
mercury arc lamp, which emits light ranging from the ultraviolet
to about 2.5 $\mu$m wavelength \cite{oriel}.  
This spectrum can be filtered to
remove light with frequencies higher than an atom's 
principle transition, so that atoms will be attracted to 
a region of high intensity.  For instance, Rb atoms have
principle transitions to the 5P$_{1/2}$ state at 794 nm and 
to the 5P$_{3/2}$ state at 780 nm, so light with wavelengths from
about 800 nm to 2.5 $\mu$m could be used.  
The confining potential $V$ depends on the 
ac polarizabilty of the atoms, $\alpha$, through \cite{Ohara99}
\be
V(\vec{r}) = -\frac{2\pi}{c} \int \alpha(\omega) S(\vec{r},\omega) \rmd\omega
\ee
where $S(\vec{r},\omega)$ 
is the spectral density of the light at position $\vec{r}$.  
At sufficiently
low frequencies, the polarizability is well approximated by the
static polarizability $\alpha\sub{s}$, 
approximately $4.7\times 10^{-29}$ m$^3$ for
Rb \cite{Miller77}.  
The dominant contribution to $\alpha\sub{s}$ is from the principle
resonances, which give a contribution
\be
\alpha\sub{r}(\omega) = \frac{1}{2} \frac{\Gamma c^3}{\omega_{1/2}^3}\left(
\frac{\omega_{1/2}}{\omega_{1/2}^2-\omega^2} + 
\frac{2\omega_{3/2}}{\omega_{3/2}^2-\omega^2} \right)
\ee
where $\Gamma = 2\pi\times 5.9$ MHz is the transition linewidth 
of the P$_{1/2}$ transition and
$\omega_{1/2}$ and $\omega_{3/2}$ are the respective transition frequencies. 
The difference between $\alpha\sub{s}$ and $\alpha\sub{r}(0)$ is about 15\%.
Across the whole frequency range, we therefore approximate the
polarizability by
\begin{equation} \label{eq-alpha}
\alpha(\omega) = \alpha\sub{s} -\alpha\sub{r}(0) + \alpha\sub{r}(\omega).
\end{equation}
Our approximation slightly underestimates the polarizability since
it neglects the growing resonant contributions of higher-lying states.
To bound the error, we compared to an overestimate obtained by assuming that
difference $\alpha_s-\alpha_r$ was entirely due to the next-lowest dipole
allowed transition at 420 nm.  The difference in $\alpha(\omega)$ was 
below 2\% over the relevant frequency range.

To maximize the light intensity, the arc should be imaged onto
the location of the atoms, as shown in figure~1.  The spectral
density for a single beam is then given by \cite{Born99b}
\be
\label{eq-s0}
S_0(\omega) = \frac{1}{2} \pi\theta^2 B(\omega) t(\omega)
\ee
where $\theta$ is the final focusing angle as shown, $B(\omega)$ is
the spectral radiance of the source, $t(\omega)$ accounts for the
imperfect transmission of any filters or other optical elements, and
the factor of 1/2 accounts for the beam splitter.  In the infrared region,
the Hg arc lamp spectrum is fairly flat as a function of wavelength.
For instance, the Oriel Instruments 100 W lamp model 6281 has a reported
brightness $B_\lambda$ of about $5 \times 10^5$ W sr$^{-1}$ m$^{-2}$ nm$^{-1}$
\cite{oriel}.
The spectral radiance therefore increases in the infrared,
\be
B(\omega) = \frac{2\pi c}{\omega^2} B_\lambda 
\ee
which helps to compensate for the decreasing polarizability.

The interference pattern produced by a spatially incoherent source has
the form
\be
S(\vec{r},\omega) = 2 S_0 [1 + \mu_{12} \cos k(z_1-z_2) ]
\ee
where $k = \omega/c$ and $\mu_{12}(\vec{r}_1,\vec{r}_2,\omega)$ is the 
spectral degree of coherence function.  Here $\vec{r}_1 = \vec{r} - \vec{p}_1$
is the displacement of the field point $\vec{r}$ from an arbitrary reference 
point $\vec{p}_1$ in one of the beams and $\vec{r}_2$ is the displacement 
from the corresponding reference in the other beam, illustrated by the
two coordinate systems in figure~1.  In particular, $z_1$ and $z_2$ are
the distances along the beam axes for the two paths.
The van Cittert--Zernike theorem gives the spectral degree of coherence 
\be
\label{eq-mu12}
\mu_{12} = 2 \frac{J_1(k\rho\theta)}{k\rho\theta}.
\ee
Here 
$\rho^2 = (x_1-x_2)^2 + (y_1-y_2)^2$ for $x$ and $y$ in the transverse plane.

In order to obtain constructive interference at the 
centre of the pattern, it is clear that both
light beams must encounter a similarly even or odd number of
mirror reflections so that the reflection
phase shifts cancel.  This means that the two images will have the
same parity, making it impossible to align both
the $(x_1,x_2)$ and $(y_1,y_2)$ axis pairs.  The Bessel function
in equation~(\ref{eq-mu12}) therefore contributes nontrivially.  
In the simplest situation,
the $y$ axes can be taken aligned, so that $\rho = 2x$.  (More generally,
new coordinates $(x',y')$ can always be found in which
the $y'$ axes are aligned and the $x'$ axes are opposed.)  The
spectral density is then given by
\be
\label{eq-stot}
S(\vec{r},\omega) = 2 S_0 \left(1 + \cos 2kz \frac{J_1(2kx\theta)}{kx\theta}
\right).
\ee
The resulting potential along the $x$ and $z$ axes is shown in figure~2.
\begin{figure}
\hfill\epsfig{file=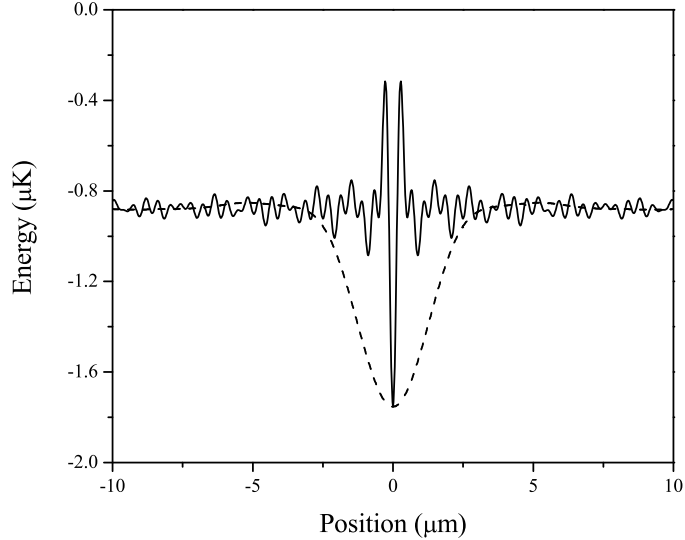,clip=,width=4in}
\caption{
Potential energy of white-light trap.  The solid trace gives the
potential along the $z$-axis and the dashed trace along the $x$-axis, as 
defined in equation (\ref{eq-stot}).  The focusing angle is $\theta = 0.15$
and the filter transmission is $t = 0.133$.
The calculation uses the measured spectral radiance curve from the
Oriel Instruments catalogue for the model 6281 lamp.  The filter
function was assumed to increase smoothly from 0 to $t$ between wavelengths
of 800 nm and 810 nm, be constant between 810 nm and 2.4 $\mu$m, and drop
smoothly to zero at 2.5 $\mu$m.  
}
\end{figure}
In addition, $S_0$ itself varies with position in a way that depends on the
details of the imaging system.  The Oriel lamp has a nominal source
size of about 250 $\mu$m, but a wide range of imaging magnifications 
is possible, depending on experimental convenience.  The final
image size $\Sigma$ gives the length scale for 
transverse variation of $S_0$, while the scale for
longitudinal variation is $\Sigma/\theta$.  We assume that both of
these scales are sufficiently large to neglect hereafter.

A single pair of interfering beams thus generates tight confinement in
one direction, looser confinement in another, and typically weak
confinement in the third.  To obtain tight confinement in all three 
dimensions, three beam pairs can be used.  These could be obtained
either from three independent lamps or by splitting the light from 
a single lamp.  Even if a single lamp is used, mutual incoherence of the
beam pairs can be assumed, since the total optical path length for beams in
different pairs is unlikely to be equal within the accuracy
required for white-light interference.  The three-dimensional potential
will thus be separable, with the potential along each axis
given by the sum of the curves in figure~2.
Figure~3 shows the result along with some of the single-particle
bound states, for light from one lamp with
$\theta = 0.15$ and $t = \frac{1}{3}\times 0.4$.  
If desired, deeper
traps can be obtained using multiple lamps and larger focusing angles.
The parameters shown support about $70$ bound states per dimension,
with two states confined in the narrow part of the potential. 
The presence of at least one tightly bound central state is robust, 
persisting down to about a factor of ten lower intensity.

\begin{figure}
\hfill\epsfig{file=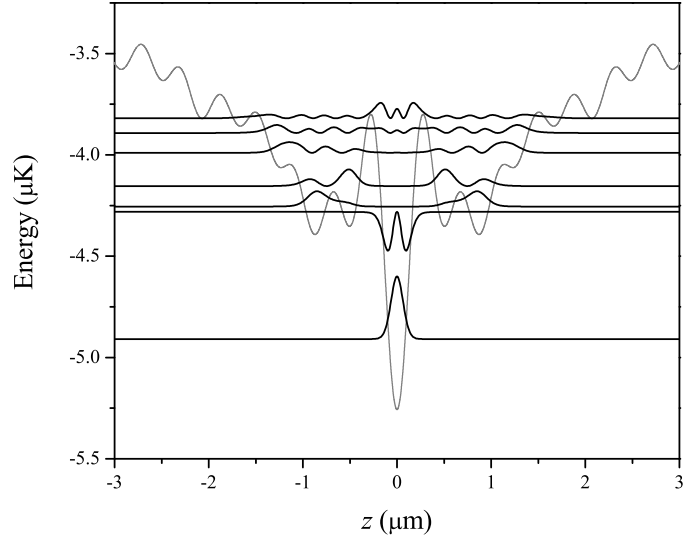,clip=,width=4in}
\caption{
Bound states of the white-light trap.  The grey curve
shows the potential along the $z$-axis of a three dimensional trap,
for Rb atoms with $\theta = 0.15$ rad and $t = 0.133$.  
The black curves give $|\psi(z)|^2$ for
the 11 lowest single-particle eigenstates, offset by the
corresponding eigenenergies.  Several of the curves are
doublets of symmetric and antisymmetric states.  The trace for the first
excited state is inverted for clarity.
}
\end{figure}

\section{Losses}

The lifetime of atoms in the white-light trap will be limited by 
various losses, some of which are novel and others
generic to tight optical traps.  
One loss mechanism is spontaneous emission, which leads to heating
and dephasing of the trapped atoms.
For both off- and on-resonant light, the scattering rate is given by the
Rayleigh formula \cite{Miller77},
\be
R_s = \frac{8\pi}{3\hbar c^4}\int S(\omega) 
\alpha^2(\omega) \omega^3 \rmd\omega.
\ee
with $\alpha$ from (\ref{eq-alpha}).  Here $S$ is the total spectral
density at the location of the atoms.  
The parameters from figure~3 give a scattering time 
$R_s^{-1} \approx 30$~s, but to achieve this rate
it is essential that light near the
atomic transitions be filtered very effectively.  For the calculation
shown, the filter function $t(\omega)$ was taken to
rise smoothly from zero at $\lambda = 800$ nm to its full value
at $\lambda = 810$ nm.  In reality, some amount of resonant light will
be present and can easily dominate the scattering rate.  
Before filtering, the spectral density near 790 nm
is $12 S_0 \approx 5\times10^{-8}$ (W/m$^2$)/Hz. 
This would give a scattering time 
of about 200 ns, so a filter attenuation
of about $10^8$ is needed.  Holographic 
notch filters can provide attenuations greater than
$10^6$ with bandwidths of about 20 nm \cite{oriel}, and
additional attenuation might
be obtained by passing the light through an optically thick Rb gas cell.
An edge filter can be used to remove light with frequencies
above the principle transitions.  

The broadband nature of the light source also permits 
resonant two-photon excitation to higher-lying excited states.  
For the range of frequencies 
applied, transitions are possible to the 6S and 4D states of Rb, both of
which lie about 20~000 cm$^{-1}$ above the 5S ground state.  Second-order
perturbation theory gives the total transition rate
for these processes to be
\newlength{\mi}
\setlength{\mi}{\mathindent}
\setlength{\mathindent}{.75in}
\be
R_2 = \sum_{b,c} \frac{\pi}{8}
\frac{ \Gamma_{ab} \Gamma_{bc}\sigma_{ab}\sigma_{bc}}{\hbar^2}
\left(\frac{1}{\omega_{bc}}-\frac{1}{\omega_{ab}}\right)
\int  
\frac{S(\omega) S(\omega_{ac}-\omega)}
{(\omega-\omega_{ab})^2(\omega-\omega_{bc})}
\rmd\omega
\ee
\setlength{\mathindent}{\mi}
where $a$ labels the ground state,
$b$ the intermediate 5P states, and $c$ the final excited 
states.  Then $\omega_{ij}$, $\sigma_{ij}$, and $\Gamma_{ij}$ are 
respectively the transition frequency, resonant scattering cross section,
and total linewidth of the
$i\rightarrow j$ transition.
The parameters used here result in a negligibly
low rate of about $5\times 10^{-3}$ s$^{-1}$.
A similar expression gives the rate for stimulated Raman transitions between
atomic ground states, which is again small compared to the Rayleigh
scattering rate.

Heating and trap loss can also arise from noise in the trap potential,
just as in laser-based optical traps.  Intensity fluctuations
with fractional noise spectrum $S_I$
lead to heating at a rate \cite{Savard97}
\be
\Gamma_\epsilon \approx \frac{\pi}{2} \Omega^2 S_I(2\Omega)
\ee
where $\Omega$ is the classical oscillation frequency, about 
$2\pi\times 15$ kHz for the parameters used above.  A 10 s heating time is then
obtained for $S_I(f = 30 \mbox{kHz})^{1/2} \approx 10^{-5}$ Hz$^{-1/2}$.  
Typical
free-running noise for research-grade arc lamps is specified to
be better than $5\times 10^{-3}$ rms in a 0.1 to 100 Hz bandwidth \cite{oriel}.
At higher frequencies, noise power is typically observed to fall as 
$1/f$ \cite{Cochran77}, leading to an expected noise density at 30 kHz 
below $10^{-6}$ Hz$^{-1/2}$.  

Difficulties can also be expected from the
phenomenon of arc wander, in which the bright spot of the arc changes
position over time \cite{oriel,Cochran77,Ebert01}.  
The time scale for this motion is again slow,
on the order of Hz, with an amplitude typically
a small fraction of the size of the arc \cite{Ebert01}.
Arc wander will not directly
change the interference pattern,
but can alter the potential strength as the 
images of the source move relative to the atoms.  
The motion is slow enough for the atoms to adiabatically follow, so
heating should be negligible.  However,
the noise introduced would be 
undesirable for tunneling experiments such as those proposed below.
One solution is to use a large demagnification ratio when imaging
the light onto the atoms.  If the geometrical image of the arc
is smaller than the aberration-limited
spot size of the system, then the relative wander of the images will be
reduced.  This does, however, lower the intensity below the geometrical
value (\ref{eq-s0}), so quantitative measurements of arc wander in a given
lamp would be a necessary step in designing a trap system.

The presence of aberrations in the imaging systems of figure~1 might
also be of concern.  
Certainly, producing a clean interference signal will require the imaging
elements of both beams in a pair to be identical to high degree.
However, the interference pattern does not in fact
depend on aberrations common to both beams, 
so long as the geometrical
image of the source is large compared to the diffraction-limited
spot size $\Sigma_{DL}$ of the imaging system \cite{Born99c}. 
Here $\Sigma_{DL} \approx 0.6 \lambda\sub{max}/\theta$ is about 10 $\mu$m. 
It is thus desirable to have a relatively large, aberration-limited image.
This is consistent with the earlier assumption of a large image,
and also beneficial because diffraction-limited 
imaging optics with the wavelength range and numerical aperture 
required would be complex and expensive.

Finally, losses may result from inelastic interactions between the trapped
atoms.  In this regard, the 
white-light trap is no different from an optical lattice with comparable
confinement strength.  The peak density for the ground state shown in 
figure~3 is $n_1 \approx 2\times 10^{14}$ cm$^{-3}$ 
per atom, and at this density,
three-body recombination occurs at a significant rate.  The
rate coefficient for condensed $^{87}$Rb atoms in the $F = 1, m_F = -1$ 
hyperfine state is 
$K \approx 6\times 10^{-30}$ cm$^6$/s \cite{Burt97}, 
defined for large numbers of
atoms $N$ with density $n(r)$ by 
\be
\frac{\rmd N}{\rmd t} = -K \int n(r)^3 \rmd ^3r.
\ee
Allowing for the loss of three atoms per collision,
counting actual triplets of atoms, and assuming a Gaussian density 
distribution, the collision rate $R_c$ is
\be
\label{eq-3body}
R\sub{c} \approx \frac{Kn_1^2}{3^{5/2}} N(N-1)(N-2),
\ee
so three atoms would last for about 10 s before recombining.

Two-body dipolar decay is suppressed for 
this hyperfine state, with an expected rate constant on
the order of $10^{-17}$ cm$^3$/s \cite{Boesten96}.  More significant will be
light-induced losses due to photoassociation (PA) of atom pairs.
Accurate estimation of the PA rate for far off-resonant light is
difficult, as it depends on Frank-Condon overlap factors between the atomic
and molecular states that are not readily available.  
In reference \cite{Miller93}, PA was observed in a laser-cooled Rb gas,
with about 100 transitions occurring in the relevant
wavelength range.
Loss coefficients ranging from $10^{-11}$ to $10^{-10}$ cm$^{3}$/s were 
measured for
an optical intensity of 10$^{10}$ W/m$^2$.  The linewidth of the
laser used in that experiment was about $10^{10}$ Hz, so the equivalent
white-light intensity is about $10^3$ W/m$^2$.  Assuming a linear
scaling with intensity \cite{McKenzie02}, this yields an estimated total
loss coefficient of about $5\times 10^{-16}$ cm$^3$/s.  This again limits the
sample lifetime to about 10 s for three atoms.  
If necessary, PA losses can
be further suppressed by increasing the short-wavelength cutoff of the
white-light.

An additional experimental issue that we consider is alignment of the
beams and interference fringe.  This might be accomplished using
the condensate itself as a target.  Even without the interference fringe,
each beam produces an optical trap deep enough to confine many atoms
(though a magnetic gradient would also be needed to support them against
gravity).
By transferring atoms to a white-light beam trap and observing their resulting
location, the positions of the white-light beam foci can be determined and
made coincident.  The location of the interference fringes 
can then be positioned
using translation stages, as illustrated in figure~1.  The stages can be roughly
positioned by direct measurement of the beam paths, but final adjustment will
likely require each stage to be finely stepped until
its fringe is located near the condensate.  When this is achieved, the
atoms will be more tightly confined, leading to changes in either 
direct images of the condensate or images of ballistic expansion.
If one beam pair is aligned at a time, 
only one dimension will be tightly confined.  
This permits a relatively large 
number of atoms to be trapped at the fringe, which would
aid the alignment process.  
Once aligned, the location of the fringes may vary
with mechanical vibrations and thermal drifts, but as long as this variation is
slow, no heating or loss should result. 

\section{Tunneling}

From these arguments, we conclude that the white-light trap is a feasible
method for tightly confining small numbers of trapped atoms for times
on the order of 10 s.  This could permit
a variety of interesting experiments.  In particular, probing
elastic interactions between the atoms is interesting for
quantum information and related applications.
The leading elastic effect is a mean-field energy shift.
For low occupation numbers, $N$ interacting atoms have a
chemical potential
\be
\label{eq-mu}
\mu \approx - E_0 + (N-1) \frac{4\pi\hbar^2a}{m} \int |\psi_0|^4 \rmd^3r
\ee
where $E_0$ is the single-particle binding energy, $\psi_0$ is the 
single-particle ground state, $m$ is the mass and 
$a$ is the scattering length.  For larger $N$, the
actual wave function must be determined from 
the nonlinear Schr\"odinger equation, which will also alter the
collisional loss rate.
Equation (\ref{eq-mu}) will be an
adequate approximation as long as the interaction energy is small compared
to the excitation energy of the first excited state.
For $^{87}$Rb, the scattering length is 5.77 nm, and 
the parameters of figure~3 give 
an excitation energy of 620 nK and interaction strength  
$g \equiv 4\pi\hbar^2 a \int d^3r |\psi_0|^4/m k_B = 32$ nK per atom.
Thus the simple expressions above should hold for $N \approx 10$ atoms or
less; interaction losses will likely limit $N$ to this regime in any case.

One way to probe the mean-field shift is to allow the atoms
to escape the trap through tunneling and to measure the
number left behind.
In an applied potential gradient $U'$, the
single-particle tunneling rate is approximately 
\be
R_T = 
\Omega \exp\left(-\frac{2}{\hbar}\int \sqrt{2m[V(z) + U' z - \mu]}
\rmd z\right)
\ee
The exponential dependence on $\mu$ provides sensitive 
energy resolution.
In gravity, for instance, the tunneling rate at the highest-lying
energy shown in figure~3 is about $7\times 10^3$~s$^{-1}$, while for the 
next highest it is $2.6\times 10^{-2}$ s$^{-1}$.  By applying
a stronger gradient, the interaction shift itself can be resolved:
in a gradient of $k_B\times$1.05 K/m, a single particle in the trap ground
state has a lifetime of 0.7 s while a second particle
escapes in 85 ms.  In this case, if the gradient is applied for
200 ms then the probability for two atoms to remain is
about 0.05, for one is 0.75, and for none is 0.20.  By measuring
these probabilities as a function of the gradient strength and
duration, the dependence of $\mu$ on 
$N$ can be experimentally determined. 
Evidently, tunneling can also be used to prepare atomic samples
with sub-Poissonian accuracy in $N$, a result useful for
many quantum information applications.

A variety of other tunneling experiments can be considered.
For instance, the white-light trap could be positioned
at the edge of a large conventionally trapped condensate.  For close spacings,
atoms can be expected to tunnel between the condensate
and the white-light trap  
when the chemical potentials of the two systems are equal.
So, as the chemical potential in the white light trap is varied
by, for instance, changing the intensity, a series of tunneling steps will
occur that are reminiscent of the Coulomb blockade effect in quantum 
dots.  This idea is explored further in Ref.~\cite{genya}.

A final interesting possibility is 
to study atoms confined in the white-light trap in 
the vicinity of a Feshbach resonance.  At an average density of
$10^{14}$ cm$^{-3}$ per atom,
the strongly interacting regime $na^3 \sim 1$ can be reached 
with $N \approx 5$ atoms if the scattering length is increased to
about 100 nm.  Enhancements of this size and larger have been
readily observed in $^{85}$Rb \cite{Cornish00}.  
In this regime, calculation of
the energies, excitations, and loss rates of the system is not trivial,
so the white-light trap could provide a testing ground for a
variety of many-body quantum techniques. 

\section{Conclusions}

In conclusion we have outlined a proposal for trapping ultracold atoms with a 
white-light interference fringe.  Using a relatively modest apparatus,
this technique can provide a single trap with oscillation frequencies of
tens of kHz and total depth of a few $\mu$K.  We estimate the lifetime of
small numbers of atoms in the trap to be of order 10 s.  
Although this paper has considered the
specific example of Rb atoms, the white-light method is
considerably more general.  Any atomic species should
be trappable, and even a variety of molecules if the absorption of
infrared light can be avoided.  An important limitation, of course,
is the initial need for an ultracold source.

We believe the white-light technique to be well-suited for studying interaction
effects in few-atom systems.  Tunneling processes can be used to
probe the energy spectrum of the atoms with nearly 
single-particle resolution.  This would provide a 
unique test of theory for weakly interacting bosons and offers the 
opportunity to observe correlation effects in more strongly
interacting systems.  In some respects,
the trap would serve as a bosonic version of a quantum dot, in
which interactions play an important, nontrivial, but 
calculable role in determining the state of the system.

A variety of applications relevant to quantum information
can be foreseen.  One interesting possibility is the use of
the white-light trap as a very small optical tweezer, capable of
positioning atoms at specific sites within a conventional optical
lattice.  Although translating the white-light trap may be difficult,
in this case the lattice itself could be translated instead.  
This could overcome one of the chief obstacles to quantum computing
in an optical lattice, the difficulty of individual addressing.
We are enthusiastic about such applications, and believe the
white-light trap to be worth further study.

\ack

We are grateful for helpful discussions with P Arnold,
E B Kolomeisky, R R Jones and T F Gallagher, and we thank
P Berger and P Uttayarat for their assistance.  This work was supported by the
Alfred P Sloan foundation.  B Deissler acknowledges support 
from the German Academic Exchange Service (DAAD).


\section*{References}

\end{document}